\documentstyle[11pt,newpasp]{article}

\def\etal{{\it et al.\  }}
\begin{document}

\title{The visible environment of gas-accreting galaxies}
\author{D. Bettoni(1), G. Galletta(2) and F. Prada (3)}
\affil{(1)Osservatorio Astronomico di Padova \\
(2)Dipartimento di Astronomia, Universit\'a di Padova \\
(3)Centro Astron\'omico Hispano-Alem\'an, Almer\'\i a, Spain}

\begin{abstract}

We studied the environment of a sample of galaxies in which 
the presence of polar rings or the presence of gas- and stars- 
counterrotation is observed. These galaxies are believed to have 
accreted this material, now in peculiar motion, from their environment.

The variable considered here are the number of possible companions in the
field, down to an apparent magnitude 22, their size, their
concentration around the accreting galaxy, and a set of related
parameters. A control sample of 'normal' galaxies has also been
studied.  From Kolgomorov-Smirnov tests of the considered variables,
no significant differences have been found between the population of
objects around the our accreting galaxies and the control sample.

These results seems to give support to the models suggesting a long 
formation time for the acquisition process, starting from a diffuse gas
instead of fagocitation of small satellite. 

\end{abstract}


\section{Introduction}

The process of galaxy formation seems to continue by means of
accretion of matter from the exteriors, at epochs much later than that
in which a galaxy has reached a stable configuration.  Peculiar
morphologies and kinematical configurations such as inclined, polar or
warped rings, observed in about a hundredth of galaxies (Bertola \&
Galletta 1978, Whitmore \etal 1990) are widely attributed to a
'second event' in the history of galaxies. The same origin is
attributed to other peculiarities, such as counterrotation, a
phenomenon characterized by gas and/or stars rotating in opposite
direction to the most part of the galaxy (Bettoni 1984, Galletta 1987,
Ciri \etal 1995).  These configurations, defined by two distinct and
eventually opposite spins coexisting in the same galaxies is not
expected in the conventional picture of galaxy formation, driven by a
sequential condensation under the effect of the gravitational force.

If there are few doubts about the external origin of the gas or stars in 
inclined or retrograde orbits, several different hypotheses have been 
presented about the origin of this matter.

On one side, the new matter has been attributed to the acquisition and
engulfment of satellite galaxies, destroyed after the merging (Quinn
1991, Rix \& Katz 1991).  But the presence of many systems with a
large amount of accreted matter has put back this hypothesis, being a
dwarf galaxy too small to justify the large detected masses in a
single event of cannibalism (Richter \etal 1994, Galletta \etal 1997).
In addition, the presence of accreted matter in spiral galaxies also
has added the difficulty to explain the accretion of large matter in a
single event without heat or destroy the disk (Barnes 1992, Rix \etal
1992, Quinn \etal 1993).

On the other side, the possibility has been studied that a large quantity of 
gas may be accreted by means of a progressive infall of diffuse matter 
(Ostriker \& Binney 1989, Thakar \& Ryden 1996).

In the first hypothesis (accretion of dwarf galaxies in an environment
particularly rich) it is possible to detect some peculiarity of the
environment by means of statistical studies in visible wavelengths of
the region of sky around the gas accreting galaxies. The second
hypothesis, accretion of diffuse gas, is hard to test from the point
of view of statistical studies.

We started a study oriented to analyze statistically the objects present in
the sky around a set of accreting galaxies, separating the polar ring
galaxies from the cases of counterrotation. We present here the preliminary 
results.

\section{Data selection and analysis}

The study of the environments of these fields was based on the counts
of the objects present and on the statistical analysis of their
properties, such as the projected distance {\it r} from the
counterrotating or polar ring galaxy and the apparent diameter D of
every small, diffuse object present in the field.  The positions and
diameters of such objects were extracted from the APM archive (Irwing
\etal 1994) available at the Observatory of Edinburgh.  We extracted
data for the field -- 200 kpc wide -- of 31 polar ring galaxies
(Brocca \etal 1997) and 43 counterrotating galaxies (Galletta
1996). As control samples for each of the previous type of peculiar
galaxies, we adopted a sample of 48 galaxies without polar ring and
another one of 53 galaxies without counterrotation. The latter sample
has been chosen looking at the published rotation curves of gas and
stars of all these stellar systems, starting from the catalog
published by Prugniel \etal (1998).

According to similar studies (Heckman \etal 1985, Fuentes-Wiliams \&
Stoke 1988) we defined for each field around the peculiar galaxy the
following density parameters:

$$\rho_{ij}= \sum_{k} r^{-1}_{k} D^{-1}_{k}$$

\noindent
where (i,j) could assume the values (0,1),(2,2) and (3,2.4). From the
above formula, $\rho_{00}$ represent the number of neighboring
galaxies, $\rho_{01}$ is the number weighted by the relative size,
$\rho_{10}$ is weighted by proximity and $\rho_{11}$ is weighted by
size and proximity. The parameter $\rho_{22}$ is proportional to the 
gravitational force exerted by the surrounding galaxies
on the central object, while $\rho_{3,2.4}$ is proportional to the tidal
interaction between the surrounding galaxies and the central one. The
last two parameters amplify the effects present in the parameter 
$\rho_{11}$ and generally a variation of one of the above independent
parameters  $\rho_{00}$, $\rho_{01}$ and $\rho_{10}$ induces changes
in the connected parameters.

Finally, after determining the $\rho_{ij}$ parameters for the two samples,
a Kolmogorov -Smirnov test has been applied to the analyzed and the control 
samples by means of a Fortran program that utilizes the IMSL library routine.

The results of this analysis are reported in the following Table.

\begin{table}
\caption[table1]{Summary of Kolgomorov-Smirnov tests. 
D$_\alpha$ is the maximum difference observed between the two distributions,
while SL is the percentage significance level at which the two distributions 
compared are different.}
\begin{center}
\begin{tabular}{lrr}
 & & \\
\hline
 & & \\
\multicolumn{3}{c}{Polar ring vs. non-polar } \\
\multicolumn{3}{c}{ ring galaxies} \\
 & & \\
\hline
 & & \\
Parameter & D$_\alpha$ & SL \\
          &            & (\%) \\
\hline
 & & \\
$\rho_{00}$ & 0.14 & 10\% \\
$\rho_{01}$ & 0.09 & 1\% \\
$\rho_{10}$ & 0.18 & 32\% \\
$\rho_{11}$ & 0.17  & 28\% \\
$\rho_{22}$ & 0.19 & 37\% \\
$\rho_{3,2.4}$ & 0.19  & 41\% \\
 & & \\
\hline
 & & \\
\multicolumn{3}{c}{Galaxies with counterrotation vs. } \\
\multicolumn{3}{c}{galaxies with co-rotation } \\

 & & \\
\hline
 & & \\
$\rho_{00}$ & 0.18 & 62\% \\
$\rho_{01}$ & 0.24 & 89\% \\
$\rho_{10}$ & 0.21 & 77\% \\
$\rho_{11}$ & 0.23  & 86\% \\
$\rho_{22}$ & 0.16 & 50\% \\
$\rho_{3,2.4}$ & 0.15  & 43\% \\
 & & \\
\hline
\end{tabular}
\end{center}

\end{table}

\section{Discussion}

The study of the environment based on the analysis of the $\rho_{i,j}$
parameters indicates that there are no significant, {\it statistical}
differences in the optical environment of accreting galaxies
(polar ring or systems with counterrotation) with respect to that of
normal galaxies. None of the significance levels for the parameters is
above 89\%.  This result is achieved also separating the sample
according to the type of the central galaxy (Elliptical, S0 or Spiral)
or according to the kind of counterrotation present (gas vs. stars or
stars vs. stars).

Our data suggest that the event of accretion generating the polar
ring or the counterrotation has not left clear, detectable traces in
the present.  According to the models existing in the literature and
as said in the introduction, two hypotheses may hold: 1) The accreted
matter is not related to the presence of satellites and derives from a
slow, non-traumatic gas infall; 2) The polar rings or the
counterrotation was generated by a mass transfer from a companion
galaxy or by a satellite ingestion, which is happened in a remote
epoch leaving traces no longer visible inside and around the galaxies.

\acknowledgments
This work has been partially supported by the grant 'Formazione ed
Evoluzione delle Galassie' Fondi ex 40\%
of the Italian Ministry of University and Scientific and Technologic 
Research (MURST).

\end{document}